\documentstyle[11pt,epsf,psfig]{article}
\def\Ex#1{\langle#1\rangle}
\def\qed{\nobreak\kern 1em \vrule height .5em width .5em depth 0em}
\def\vbar{\mathchoice{\vrule height6.3ptdepth-.5ptwidth.8pt\kern-.8pt}
   {\vrule height6.3ptdepth-.5ptwidth.8pt\kern-.8pt}
   {\vrule height4.1ptdepth-.35ptwidth.6pt\kern-.6pt}
   {\vrule height3.1ptdepth-.25ptwidth.5pt\kern-.5pt}}

\let\wt\widetilde
\def\date
   {\noindent Date: \today \par
    \medskip}

\def\bra#1{\langle #1 |} 
\def\ket#1{| #1 \rangle}

\def\G#1{\Gamma(#1)}

\def\t{\int_0^tdt}
\def\p#1{\frac{\partial}{\partial#1}}

\def\dr{\int_{z>0} d^dr}

\def\e{\epsilon}
\def\f{\phi}

\def\s{\overline{\phi}}
\def\lt#1{\widetilde{#1}}

\def\dra{\int d^dr}
\def\S{{\cal S}}
\def\G{{\cal G}}
\newcount\subno      
\newcount\subsubno   
\newcount\appno      
\newcount\appeqno    
\newcount\tableno    
\newcount\figureno   
\def\nsection#1
   {\vskip0pt plus.1\vsize\penalty-250
    \vskip0pt plus-.1\vsize\vskip24pt plus12pt minus6pt
    \subno=0 \subsubno=0
    \addtocounter{section}{1}
\renewcommand{\thesection}{\Roman{section}}
  {\small  \noindent {\bf \thesection. #1\par}}
    \noindent}
\def\nsubsecnn#1
   {\vskip-\lastskip
    \vskip12pt plus12pt minus6pt
    {\noindent {\bf #1\par}}
    \noindent}
\begin{document}
%
\ \\[12mm]
\begin{center}
    {\bf REACTION DIFFUSION AND BALLISTIC ANNIHILATION\\ 
	NEAR AN IMPENETRABLE BOUNDARY}

\end{center}
\begin{center}
\normalsize
Y. Kafri and M. J. E. Richardson\\
{ \it Department of Physics of Complex Systems, The Weizmann Institute of Science\\ 
Rehovot 76100 Israel.\\[3mm] }
\end{center}
\noindent {\bf Abstract:} 
The behavior of the single-species reaction process $A+A\rightarrow O$ is examined near an impenetrable boundary, representing the flask containing the reactants. Two types of dynamics are considered for the reactants: diffusive and ballistic propagation. It is shown that the effect of the boundary is quite different in both cases: diffusion-reaction leads to a density excess, whereas ballistic annihilation exhibits a density deficit, and in both cases the effect is not localized at the boundary but penetrates into the system. The field-theoretic renormalization group is used to obtain the universal properties of the density excess in two dimensions and below for the reaction-diffusion system. In one dimension the excess decays with the same exponent as the bulk and is found by an exact solution. In two dimensions the excess is marginally less relevant than the bulk decay and the density profile is again found exactly for late times from the RG-improved field theory. The results obtained for the diffusive case are relevant for Mg$^{2+}$ or Cd$^{2+}$ doping in the TMMC crystal's exciton coalescence process and also imply a surprising result for the dynamic magnetization in the critical one-dimensional Ising model with a fixed spin. For the case of ballistic reactants, a model is introduced and solved exactly in one dimension. The density-deficit profile is obtained, as is the density of left and right moving reactants near the impenetrable boundary.
\\[5mm]
\date 
\rule{7cm}{0.2mm}
\begin{flushleft}
\parbox[t]{3.5cm}{\bf Key words:}  non-equilibrium statistical mechanics, reaction process, ballistic annihilation, renormalization group,exactly solved models, field theory
\parbox[t]{12.5cm}{ }
\\[2mm]
\parbox[t]{3.5cm}{\bf PACS numbers:} 02.10.Eb, 05.40.+j, 05.70.Ln, 64.60.Cn
\\[2mm]
\end{flushleft}
\normalsize
\thispagestyle{empty}
\mbox{}
\pagestyle{plain}

\newpage

\setcounter{page}{1}
\pagestyle{plain}
\setcounter{equation}{0}

\normalsize
\thispagestyle{empty}
\mbox{}
\pagestyle{plain}
\setcounter{page}{1}
\setcounter{equation}{0}

\noindent{\Large{\bf Introduction}}\\[5mm]
Systems of reacting and coalescing particles, whether chemical or more exotic, are excellent examples of statistical systems far from equilibrium. Such processes are widespread throughout nature and cover a broad range of physical phenomena. Though the microscopic details of specific reaction systems may vary enormously, it is often the case that the global behavior of such processes is determined by only a few of its properties: the number of particles that interact locally in a single reaction and the dynamics of a single reactants' motion. Many of the simple reaction schemes have been extensively studied, for example the first-order reaction processes $A+A\rightarrow O$ or $A+A\rightarrow A$. Also, much is known of the more complicated two-species process $A+B\rightarrow O$ where the formation of reaction fronts \cite{GALFI} and spontaneous segregation of species \cite{TW} can occur. These reaction processes have been analyzed for the case of reactants that diffuse and also that have ballistic motion. The former is the appropriate choice for systems where the mean-free path is much less than the inter-reactant distance, and the latter is appropriate for the contrary case, for example gas-phase reactions. Of course, given that the inter-reactant distance is generally decreasing as the reaction process continues it can occur that in some physical systems a cross-over will exist from ballistic to diffusive behavior \cite{BUT}.

The majority of studies have considered reaction processes that occur in a translationally invariant system, for example in systems with periodic boundary conditions or of infinite extent. In this paper we analyze the $A+A\rightarrow O$ reaction near a fixed, impenetrable boundary. Such a boundary represents the vessel containing the reactants, though as will be discussed, it can also represent other phenomena in esoteric reaction systems.

In section 1 we describe the case of diffusive reactants near a boundary and show that a {\it density excess} is formed which extends into the system. 
A concise account of these results has been presented elsewhere \cite{RK}, but without a detailed description of the method. The late-time forms of this density excess are found exactly in one and two dimensions and we discuss a surprising implication for the dynamic magnetization of the critical one-dimensional Ising model. As a counter-point to the diffusive case, in section 2 we consider the case of ballistic annihilation near an impenetrable boundary and show that it has the opposite effect: a {\it density deficit} is formed at the boundary. A model for the ballistic process is introduced and solved exactly with the density profiles of left and right moving reactants obtained. Finally, we close with a discussion of the results obtained and open questions.

\section{Reaction-diffusion near a boundary}
In this section the effect of an impenetrable boundary on the dynamics of a
 single-species reaction process with {\it diffusive} reactants is studied. 
 This universality class comprises the annihilating random walk
 $A+A \rightarrow O$, coalescing random walk $A+A \rightarrow A$ and 
any combination thereof \cite{PELITI}. Though the results below will be explicitly derived for the annihilating random walk, they are identical for every member of the universality class, albeit with a trivially altered pre-factor. As well as chemical processes, the dynamics of single-species reactions systems are exhibited in a broad range of 
physical phenomena. For example, the annihilating random walk models the
 domain coarsening of the one-dimensional critical Ising model \cite{ISING}. Starting from an initially uncorrelated random state at zero temperature, the Ising system evolves by Glauber dynamics. The domains increase in size by the process of domain wall annihilation and an isolated domain wall performs a random walk (in zero applied field). Such a domain wall can therefore be thought of as a reactant in the single-species $A+A\rightarrow O$ reaction. (A similar mapping holds between the coarsening of the $\infty$-state Potts model and the $A+A\rightarrow A$ system). A second example is the exciton coalescence reaction $A+A\rightarrow A$ which is seen experimentally in the TMMC crystal \cite{TMMC,PRIV}. The non-trivial decay exponent predicted theoretically for the laser-induced electronic excitations is seen over many orders of magnitude. The boundary conditions to be studied in this section are realized in the TMMC crystal with doping by Mg$^{2+}$ and Cd$^{2+}$ ions that act as perfect reflectors for the annihilating excitons \cite{MGCD}.
 
We first review the relevant results for the reaction process in a translationally invariant unbounded system, which will be referred to as the {\it bulk} case. Following that the model will be defined and the basic results stated. A detailed description of the method will then be given. The dynamics will be written in terms of a master equation and then mapped to a field-theoretic representation. The universal properties of the density excess will be obtained and the asymptotically-exact density profile found in two dimensions. Motivated by the mapping to the Ising system, a model will introduced and solved exactly in one dimension. The section is closed with a summary of results obtained.
%
%
%
%
%
%
%
%
%
\nsubsecnn{Summary of known results for the unbounded system}
A wealth of research exists on the single-species $A+A\rightarrow O$ process in unbounded systems \cite{TW}, \cite{KANGRED}--\cite{DOI}. There are many realizations of the dynamics, but the resulting late-time behaviors show a strong degree of universality and are independent of the details of particular models. The annihilating random walk describes the process whereby diffusing $A$ particles, moving in a $d$-dimensional space, may react pairwise on contact, at a rate $\lambda$. The principal quantity of interest is the decay of the density $\varrho$ of reactants as a function of time $t$. A starting point in the analysis of such systems is the mean-field approach. This corresponds to writing a self-consistent equation for the
 average reactant density $\overline{\varrho}$, thereby ignoring all spatial correlations
\begin{eqnarray}
\frac{\partial \overline{\varrho}}{\partial t}=D\nabla^2\overline{\varrho}-2\lambda\overline{\varrho}^2,&\mbox{ with the late time result }&\overline{\varrho}\simeq\frac{1}{2\lambda t} \label{mf}
\end{eqnarray}
where we have introduced the diffusion constant $D$. Neglecting the effect of
 correlations is equivalent to assuming that the reactants remain well mixed
 throughout the reaction process. However, due to the statistics of random
 walks in two dimensions and below, simple diffusion of particles itself is
 not sufficiently fast to maintain a well-mixed state. Reactants that are
 close together react fast leaving reactants that are more widely separated,
 i.e. anti-correlated in space. For this reason, the mean-field result
 (\ref{mf}) loses its validity below two dimensions. Also, in these lower
 spatial dimensions the fact that random walks are recurrent means that
 reactants come
 into contact many times, each time providing an opportunity for a reaction
 to take place. This implies that even a small reaction rate $\lambda$ does
 not limit the global rate of reaction. In two dimensions and below the
 reaction becomes {\it diffusion limited} and is independent of the
 parameter $\lambda$. The correct forms for the density in two dimensions
 and below are given in Table \ref{bulktable} at the end of this section.
%
%
%
%
%
%
%
%
%
%
\nsubsecnn{Reactions near a boundary - new results}
We now define the model that will be analyzed throughout the remainder of
 this section. The dynamics of the $A$ particles are the same as for
 the unbounded case except that now the geometry is restricted by the imposition of a hyperplane boundary, confining all particles to the positive half-volume. To be specific, the model is
 defined on a hypercubic, $d$-dimensional lattice which for convenience has a 
lattice spacing of unity. This $d$-dimensional lattice is infinite in $d-1$
 transverse dimensions and semi-infinite, sites $1,2,\cdots,\infty$ in what
 will be called the $z$ direction. Initially, the lattice
 is filled with an average density $\varrho_0$ of $A$
 particles that have two components to their behavior, diffusive motion and
 mutual annihilation. They can diffuse randomly throughout the lattice by hopping at a
 rate $D$ to any neighboring site, as long as the restriction $z\geq1$ is maintained. The diffusion is 
independent for each particle and hence multiple occupancy of a lattice site
 is allowed, leading to bosonic statistics. However, if there are $n\geq2$
 particles on a site, a reaction can occur there
 with a rate $\lambda n(n-1)$ reducing $n$ by 2. 

In the same way as for the bulk case, the dynamics of the model can
 be approximated by a mean-field description of the same form as
 equation (\ref{mf}) but with the added restriction of zero current at the
 boundary. This amounts to enforcing a vanishing density-gradient at $z=0$ which is a
 restriction compatible with the unbounded solution. Hence, the mean-field
 equation asserts that the boundary has {\it no effect}. Nevertheless, as
 was shown above mean-field cannot be relied upon in two dimensions and
 below because of the importance of correlations. In fact, even a simple
 argument shows that a {\it density excess} develops near the boundary in
 low dimensions. Consider the dynamics of the model in one dimension, up to
 a time $t$ and far from
 the influence of the wall. Since random walks are recurrent in these
 dimensions, in the bulk most particles
 within a diffusion length $l_b\sim\sqrt{2Dt}$ will have
 interacted
 and annihilated. This leads to a density in the bulk of the system of
 $\varrho_b\simeq l_b^{-1}= c_b/\sqrt{t}$. Close to the wall, 
within a distance of the order of $\sqrt{Dt}$, the diffusion length is
 smaller. The density near the wall is then $\varrho_w\simeq l_w^{-1}= c_w/\sqrt{t}$, with
 $c_w\!>\!c_b$.
 As $c_w=c_w(z^2/Dt)$ the argument implies that there is a density excess
 near
 the boundary which propagates into the system diffusively.

This is a qualitative argument, however we will show analytically that in two dimensions and below this is indeed the case: the reactant density has the form of a
 constant background density $\varrho_B$ given by the well-known unbounded results in Table \ref{bulktable}, and a fluctuation-induced density excess $\varrho_E$
\begin{eqnarray}
\varrho(z,t)&=&\varrho_B(t)+\varrho_E(z,t). \label{den2}
\end{eqnarray}
As will be demonstrated, the density excess has the universal dimension-dependent form
\begin{eqnarray}
\varrho_E(z,t)&=&\frac{1}{(8\pi Dt)^{d/2}}f_d\left(\frac{z^2}{2Dt}\right).
\label{ex}
\end{eqnarray}
In the following two sub-sections we use both the field-theoretic
 renormalization group and an exact solution to determine the asymptotic forms of the scaling functions $f_2$ and $f_1$ for two and one dimensions respectively. A reasonable amount of technical details
 are given, but all the important results derived can be found
 in the summary at the end of this section.
%
%
%
%
%
%
%
%
%
%
\subsection{Two dimensions and below}
In the following we will examine the late-time scaling behavior of
 the lattice model described above. First, the dynamics will be written as a bosonic master equation and mapped to a continuum field theory via the coherent-state formalism. The theory is then regularized and the renormalization group applied to obtain the universal properties and non-perturbative results valid for late time in dimensions two and below. The method is standard
 \cite{DOI,PELITI2,MATGLA}, but is technically more complicated than the
 unbounded case \cite{PELITI,BPL} due to the lack of translational invariance
 that necessitates calculations in real space. 
%
%
%
%
%
%
%
%
%
%
\nsubsecnn{The field theory}
The first step in the field-theoretic calculation is to write the dynamics of the
 model in terms of a {\it master equation}. This describes the flow of
 probability between microstates of the system defined by the set of occupation variables $\{n_i\}$ where $n_i$ is the number of particles
 on site $i$. Due to the bosonic statistics of
 the $A$ particles it is convenient to write the master equation in a second-quantized
 form. Each configuration, $\{n_i\}$ is assigned a vector in a
 bosonic Fock space $\ket{\lbrace n_i \rbrace}$, with $\ket{0}$ denoting the no-particle state. Using $a_k$ and $a^\dagger_k$ as the bosonic operators for site $k$ the state ket of the system $\ket{P(t)}$ is written 
$$
\ket{P(t)}=\sum_{\{n\}}P(\{n_j\}:t)\prod_i(a_i^{\dagger})^{n_i}\ket{0}.
$$
Here $P(\{n_j\}:t)$ is the probability of being in a particular microstate
 $\{n_j\}$ at a time $t$, the sum is over all microstates, and the product is over
 all lattice sites. The master
 equation for the model can
 now be written in the form $\partial_t|P\!>=-{\cal H}|P\!>$ where the evolution operator ${\cal H}$ is given by 
\begin{eqnarray}
{\cal H}&=& \sum_i\left[D\sum_ja^\dagger_i(a_i-a_j)-\lambda(1-(a^\dagger_i)^2)a_i^2\right]. \nonumber
\end{eqnarray}
The sum $i$ is over all lattice sites, and the sum $j$ is over all of site $i$'s
 neighbors, with the condition that both sums are restricted to the half-space. We now introduce two objects, the projection state $\bra{}=\bra{}a^\dagger$ and the initial state $\ket{\varrho_0}$ with average particle density $\varrho_0$, that allow observables such as the density $\varrho_k(t)$ to be written
\begin{eqnarray}
\bra{}=\bra{0}e^{{\sum_j a_j}}&\mbox{ and }&\ket{\varrho_0}=e^{\varrho_0\sum_j{(a^\dagger}_j-1)}\ket{0}, \nonumber\\
\varrho_k(t) &=&\bra{}a_ke^{-{\cal H}t}\ket{\varrho_0}, \nonumber 
\end{eqnarray}
where the sums are again over all lattice sites $j$ with $z>0$. The field-theoretic representation is obtained using the coherent-state formalism \cite{BROWN}. The continuum of microstates of the system are now given by configurations of the field variables $\{\f_0,\s_0\}$ which are analogous to
 $\{a,a^\dagger-1\}$. Observables are written as integrations over the appropriate weighting function $\S_0(\f_0,\s_0)$ which we will call the {\it action}. For example, the density at position $r$ and time $t$ is written
\begin{eqnarray}
\varrho(r,t)&=&\int{\cal D}\f_0{\cal D}\s_0\;\;\f_0(r,t)\;\; e^{-\S_0(\f_0,\s_0)}, \label{den1} \\
{\cal S}_0(\f_0,\s_0)&=&{\cal S}_D+2\lambda\t\dr\s_0\f_0^2+\lambda\t\dr\s_0^2\f_0^2-\varrho_0\dr \s_0(0). \label{bareaction}
\end{eqnarray}
The diffusion component of the action ${\cal S}_D$ defines the propagator
\begin{eqnarray}
\G(r_f,r_i:t)&=& \int{\cal D}\f{\cal D}\s\;\;\f_0(r_f,t)\s_0(r_i,0)\;\;e^{-\S_D(\f,\s)} \nonumber \\
\G(r_f,r_i:t)&=&\left[G(z_f-z_i:t)+G(z_f+z_i:t)\right]\prod_{\{y\}}G(y_f-y_i:t), \label{prop}
\end{eqnarray}
where we have used the notation $r=\{z,y_1,y_2\cdots\}$, with the set $\{y\}$ representing the dimensions parallel to the boundary and the function $G(x:t)$ is the Gaussian
\begin{eqnarray}
G(x:t)&=&\frac{1}{\left(4\pi Dt\right)^{1/2}}\exp{\left(-\frac{x^2}{4Dt}\right)}. \nonumber
\end{eqnarray}
%
%
%
%
%
%
%
%
%
%
\nsubsecnn{Regularization of the theory}
From power counting, it is seen that the upper-critical dimension $d_c=2$ of
 the theory is unchanged from the bulk case. The theory contains divergences,
 by virtue of the continuum limit and to write the density in a
 calculationally useful form we need to replace all the bare quantities with
 renormalized ones and render the theory finite. This is done using 
dimensional
 regularization in $d=2-\e$. In fact the renormalization of the theory is
 straightforward as there is no propagator, initial-condition or field
 renormalization: only the reaction rate is renormalized. Because the
 propagator is not dressed by the interactions, Eq. (\ref{prop}) is
 the full result showing that the boundary remains effectively reflecting
 on all
 scales. In the language of surface critical phenomena, this corresponds to
 the
 {\it special transition} \cite{DIEHL} persisting at all orders and is
 different
 from the behavior frequently seen in equilibrium surface critical phenomena
 \cite{DD,DD2} and in related non-equilibrium systems \cite{DP,DDB}.

\begin{figure}
\epsfxsize 10 cm
\epsfysize 7 cm
\hspace{3.5cm}\epsfbox{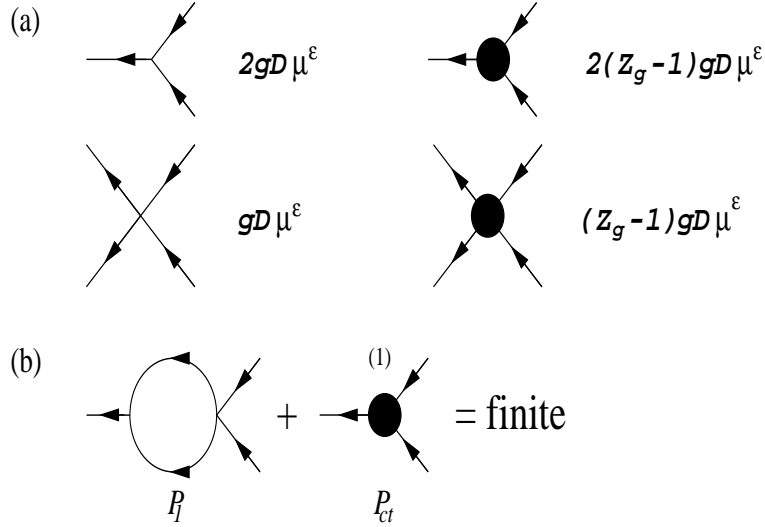}
\caption{The diagrammatic representations of (a) the interactions and counter terms (with blob) and (b) the one-loop regularization. The propagator is denoted by a straight line with an arrow signaling the time direction. In the diagramatic equation (b) $P_1$ and $P_{ct}$ are the one-loop and $Z_g^{(1)}$ counter-term contributions respectively.}\label{basicdia}
\end{figure}

The bare action ${\cal S}_0=\S+\S_{ct}$ is now rewritten in terms of the
 renormalized action ${\cal S}$ and the counter-term action ${\cal S}_{ct}$.
 We also introduce the dimensionless interaction parameter $g$ defined by
 $\lambda=Z_ggD\mu^\e$ with $\mu$ arbitrary but with dimensions of inverse 
length. Thus
\begin{eqnarray}
\S&=&\S_D+gD\mu^\e\left[2\t\dr\s\f^2+\t\dr\s^2\f^2\right]-\varrho_0\dr \s \label{action}\\
\S_{ct}&=&(Z_g-1)gD\mu^\e\left[2\t\dr\s\f^2+\t\dr\s^2\f^2\right]. \nonumber 
\end{eqnarray}
The diagrammatic representation of the two reaction terms and the counter terms are given in Fig. \ref{basicdia}(a).

The renormalization constant $Z_g=1+Z_g^{(1)}+\cdots$ can now be
 calculated up to the one-loop order. The minimal-subtraction regularization 
is done perturbatively (in the absence of the initial conditions) by demanding
 that calculated quantities at a given order of perturbation theory are finite.
As an example, consider the correlation function
$$
P=\Ex{\f(c,t)\s(a,0)\s(b,0)}=\left[P_1+P_{ct}\right]+\cdots=\mbox{finite} \nonumber 
$$
where $\Ex{\ldots}$ denotes an average of the fields over the action
 (\ref{action}) without the initial-condition term. We
 have used the notation $P_1$ to denote the contribution to this correlation
 function at the one-loop level and $P_{ct}$ is the appropriate counter-term, see Fig. \ref{basicdia}(b):
\begin{eqnarray}	
P_1&=&8gD^2\mu^{2\e}\int_0^tdt_1\dr_1\int_0^{t1}dt_2\dr_2 \nonumber \\
&&{\cal G}(c-r_1,t-t_1){\cal G}^2(r_1-r_2,t_1-t_2){\cal G}(r_2-a,t_2){\cal G}(r_2-b,t_2), \nonumber \\
P_{ct}&=&-Z^{(1)}_g 4gD\mu^{\e}\int_0^tdt_1\dr_1{\cal G}(c-r_1,t-t_1){\cal G}(r_1-a,t_1){\cal G}(r_1-b,t_1). \nonumber 
\end{eqnarray}
Using the above equations the scheme requires that \mbox{$[Z_{g}^{(1)}-2Dg\mu^\e K]$} is finite, where $K$ is defined through
\begin{eqnarray}
K&=&\int^{t_1}_0dt_2\dr_2\frac{{\cal G}(r_2-a,t_2){\cal G}(r_2-b,t_2) }{{\cal G}(r_1-a,t_1){\cal G}(r_1-b,t_1)}{\cal G}^2(r_1-r_2,t_1-t_2).  \nonumber
\end{eqnarray}
The singular part of $K$ can be extracted so that 
$$
2Dg\mu^\e K=Z_g^{(1)}=\frac{g}{2\pi\e}+O(\epsilon^0). 
$$
Hence using the renormalization constant $Z_g=1+g/(2\pi\e)$ the one-loop beta function is
\begin{eqnarray}
\beta(g)&=&\frac{g}{2\pi}(g-2\pi\e) \mbox{ in $d=2-\epsilon$}. \label{betafun} 
\end{eqnarray}
The fixed point structure is unchanged from the bulk result found in \cite{BPL}. This is understandable as physically the renormalization of $\lambda$ is connected
 to the fact that random walks are recurrent in two dimensions and below: a feature unaffected
 by the presence of a boundary. With the theory regularized, it is now possible to proceed with a perturbative calculation of the density that is valid for early times
%
%
%
%
%
%
%
%
%
\nsubsecnn{The perturbative density}
To calculate the density $\varrho$ at early times, three objects are needed: the tree-level contribution $\varrho^{(0)}$, the one-loop
 contribution $\varrho^{(1)}$ and the counter-term $\varrho^{(1)}_{ct}$  
\begin{eqnarray}
\varrho&=&\varrho^{(0)}+ \varrho^{(1)}+\varrho^{(1)}_{ct}. \label{den}
\end{eqnarray}
All these are calculated using the renormalized action with the initial conditions included
\begin{eqnarray}
\varrho(r,t)&=&\int{\cal D}\f{\cal D}\s\;\;\f(r,t)\;\; e^{-\S -\S_{ct}}. \nonumber
\end{eqnarray}
where $\S$ and $\S_{ct}$ are defined in Eq. (\ref{action}). However, as will
 be justified later, an expansion is made in the parameter $\varrho_0^{-1}$
 with only the leading term retained. This is
 equivalent to the limit of high initial density.\\

\begin{figure}
\epsfxsize 10 cm
\epsfysize 7 cm
\hspace{2cm}\epsfbox{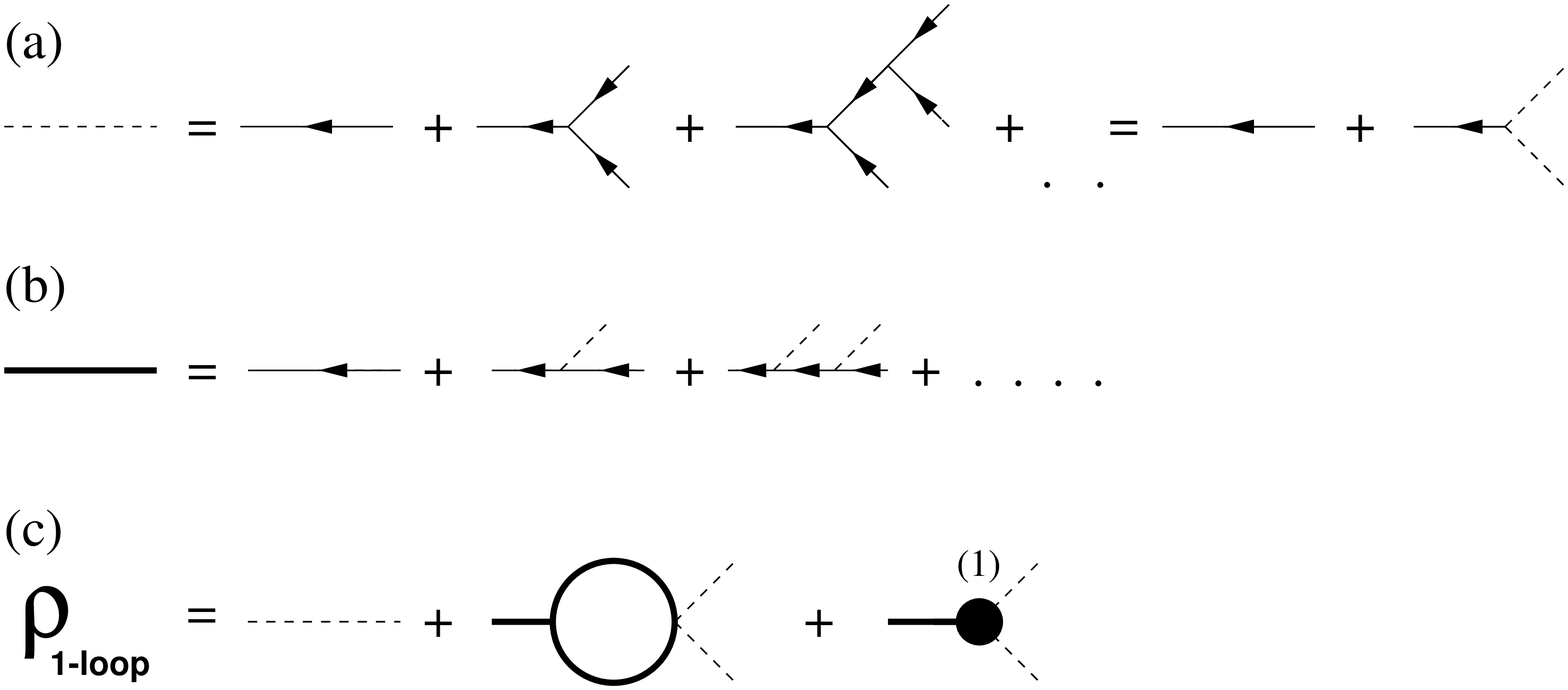}
\caption{The diagrammatic representations of (a) the tree-level density calculation, (b) the classical propagator and (c) the perturbative density to the level of one loop, which reading from left to right are $\varrho^{(0)}$, $\varrho^{(1)}$
and $\varrho^{(1)}_{ct}$ } 
\label{densitya}
\end{figure}

\noindent (i) {\it The tree-level contribution} $\varrho^{(0)}$\\
The tree-level density is obtained by the sum of all diagrams given in Fig. \ref{densitya}(a) with the sum rewritten to yield an integral equation. The result is found to be equivalent to the bulk case, as is expected from the mean-field analysis earlier.
\begin{equation}
\varrho^{(0)}=\frac{\varrho_0}{1+2\varrho_0 g\mu^\e Dt}=\frac{1}{2g\mu^\e Dt}+O(\varrho_0^{-1}).
\label{treelevel}
\end{equation}
Before proceeding to the one-loop calculation, the dressed, tree-level propagator must be calculated. Following \cite{BPL} we call this the classical propagator ${\cal G}_{C}$ and it is defined through
$$
{\cal G}_C \equiv\Ex{\f(r_1,t_1) \s(r_2,t_2)}_C
$$
 with only the tree diagrams included in the average, see
Fig. \ref{densitya}(b). This propagator is evaluated in real space to give
\begin{equation}
{\cal G}_C={\cal G}(r_1-r_2,t_1-t_2)\left(\frac{1+2 \varrho_0 g\mu^\e Dt_2}{1+2 \varrho_0 g\mu^\e Dt_1}\right)^2={\cal G}(r_1-r_2,t_1-t_2)\left(\frac{t_2}{t_1}\right)^2+O(\varrho^{-1}_0).
\label{cprop}
\end{equation}\\

\noindent (ii) {\it The one-loop contribution $\varrho^{(1)}$}\\ 
Using the classical propagator the one-loop contribution takes a simple form, see 
Fig. \ref{densitya}(c),
\begin{eqnarray}
\varrho^{(1)}&=&4D^2\mu^{2\e}g^2\int_0^tdt_1\dr_1\int_0^{t_1}dt_2\dr_2{\cal G}_{C}(r-r_1,t-t_1){\cal G}^2_{C}(r_1-r_2,t_1-t_2)\left( \varrho^{(0)}(t_2) \right) ^2. \nonumber
\end{eqnarray}
To simplify the notation we introduce $\G_B(r,t)$, the bulk propagator, defined through
\begin{eqnarray}
\G_B(r_f,r_i:t)&=&G(z_f-z_i:t)\prod_{\{y\}}G(y_f-y_i:t).
\end{eqnarray}
In the large $\varrho_0$ limit the one-loop contribution then takes the form 
\begin{eqnarray}
\varrho^{(1)}&=&\frac{1}{t^2}\int_0^tdt_1\int_0^{t_1}dt_2\left(\frac{t_2}{t_1}\right)^2(I_B+I_E), \mbox{ with}\nonumber \\
I_B&=&\dra_1\dra_2\G_B(r-r_1,t-t_1)\G_B^2(r_1-r_2,t_1-t_2)  \nonumber \\
I_E&=& \dra_1\dra_2\G_B(r-r_1,t-t_1)\G_B(r_1-r_2,t_1-t_2)\G_B(r_1-r_2',t_1-t_2).  \nonumber 
\end{eqnarray}
Note that the half-volume integrations have been exchanged by integrations over the volume. The integral has been separated into two parts $I_B$ and $I_E$ corresponding to the bulk and boundary/excess contributions. The notation $r_2'$ denotes a set of coordinates in which the $z$ 
component is minus that of $r_2$, but all other coordinates are unchanged. 
We consider these two quantities $I_B$ and $I_E$ separately.

Firstly, the bulk integral $I_B$ is performed and gives the following one-loop contribution to the perturbative density
\begin{eqnarray}
I_B&=&\left(\frac{1}{2(2\pi D(t_1-t_2))^{1/2}}\right)^d \nonumber \\
\varrho^{(1)}_B&=&\frac{1}{(Dt)^{d/2}}\left(\frac{1}{4\pi\e}+\frac{1}{16\pi}(2\log(8\pi)-5)\right)+O(\e). \nonumber
\end{eqnarray}
This result is the same as that found in \cite{BPL} and provides a uniform background density.

Secondly the excess contribution $I_E$ is considered. The
 integrations over the $d-1$ dimensions parallel to the surface yield the result as the integral $I_B$ (with the power $d$ replaced by $d-1$). However, 
the integration over the direction perpendicular to the wall yields new a $z$ dependent result. Combining these results one obtains for $I_E$
\begin{eqnarray}
I_E&=&\frac{1}{(8\pi D)^{d/2}}  \frac{1} {(t_1-t_2)^{(d-1)/2}(2t-t_1-t_2)^{1/2}}\exp\left(-\frac{z^2}{2D(2t-t_1-t_2)}\right). \nonumber
\end{eqnarray}
Using this we obtain the contribution to $\varrho^{(1)}$,
\begin{eqnarray}
\varrho^{(1)}_E&=&\frac{1}{(Dt)^{d/2}} \left(\frac{1}{8\pi}\right)f_2\left(\frac{z^2}{2Dt}\right)+O(\e),\mbox{ where} \nonumber \\
f_2\left(\frac{z^2}{2Dt}\right)&=&\int_0^1 ds\int_0^sdq \left(\frac{q}{s}\right)^2 \frac{\exp\left(-\frac{z^2}{2Dt}\frac{1}{(2-s-q)}\right)} {(s-q)^{1/2}(2-s-q)^{1/2}}.  \label{2d}
\end{eqnarray}
We were unable to evaluate this double integral, except at $z^2/Dt=0$ where the result $f_2(0)=3/2+\pi-3 \pi^2/8$ was found. However, it has the asymptotic form $f_2(\xi^2) \sim \exp(-\xi^2/2)/\xi^3$ which is valid for $\xi\gg1$.\\
 
\noindent (iii) {\it The counter term} $\varrho^{(1)}_{ct}$\\
As before in terms of
the classical propagator this contribution takes the simple form, see Fig. \ref{densitya}(c).
\begin{eqnarray}
\varrho^{(1)}_{ct}&=&-Z_{g}^{(1)}f\mu^\e D\int_0^t dt_1 \dr_1 {\cal G}_C(r-r_1,t-t_1) \left(\varrho^{(0)}(t_1)\right)^2. \nonumber 
\end{eqnarray}
In the large $\varrho_0$ limit this is evaluated in a manner similar to $\varrho^{(0)}$ to give $\varrho^{(1)}_{ct}=-Z_{g}^{(1)}/2Dtg\mu^\e$.\\ 

All the required perturbative results have now been obtained up to the order of one loop. These tree-level and one-loop results are combined and rewritten as in equation (\ref{den2}) as the sum of a spatially independent, background density $\varrho_B$ and a spatially dependent, excess density $\varrho_E$
\begin{eqnarray}
\varrho_B(t)&=&\frac{1}{2gDt\mu^\e}+\frac{1}{(Dt)^{d/2}}\left[\frac{1}{4\pi\e}+\frac{2\log(8\pi)-5}{16\pi}\right]-\frac{1}{4\pi\e Dt\mu^\e}+O(g) \label{PB} \\
\varrho_E(t)&=& \frac{1}{8 \pi (Dt)^{d/2}}f_2\left(\frac{z^2}{2Dt}\right)+O(g),\label{PE}
\end{eqnarray}
where the results here are correct at the level of one-loop. In the next section this perturbative density will be used to obtain the late-time results by the use of the renormalization group. 
%
%
%
%
%
%
%
%
%
%
\nsubsecnn{The flow equations}
A Callan-Symanzik equation for the density can be obtained by noting that any observable must be independent from the arbitrary $\mu$. Using this fact and dimensional analysis allows one to write two partial differential equations which can be combined to give
\begin{eqnarray}
&&\left[\p{\log z}+2\p{\log Dt}-d\p{\log\varrho_0}+\beta(g)\p{g}+d\right]\lt{\varrho}(z,Dt,\varrho_0,g,\mu)=0 \label{flow} \\
&&\left[\frac{d}{d\log s}+d\right]\lt{\varrho}=0, \nonumber
\end{eqnarray}
where we have rewritten the sum of partial derivatives as a total derivative with respect to a scaling variable $s$. Following the notation in \cite{BPL} we write $\lt{X}=X(s)$ for early-time quantities and $X=X(1)$ for late-time quantities,
\begin{equation}
s^d\lt{\varrho}=\varrho,\hspace{.5cm}\lt{z}=sz,\hspace{.5cm}\lt{t}=s^2t,\hspace{.5cm}\lt{\varrho_0}=s^{-d}\varrho_0,\hspace{.5cm}\lt{g}=\int\beta(g)d\log(s), \nonumber
\end{equation}
and obtain the exact relation between a density with arguments $\{z,Dt,\varrho_0,g,\mu\}$ to
 a density with $\{\lt{z},D\lt{t},\lt{\varrho_0},\lt{g},\mu\}$
\begin{equation}
\varrho(z,Dt,\varrho_0,g,\mu)=\left(\frac{D\lt{t}}{Dt}\right)^{d/2}\lt{\varrho}(\lt{z},D\lt{t},\lt{\varrho_0},\lt{g},\mu).
\label{denflow}
\end{equation}
Next we will take the limit of large $t/\lt{t}$ ($s\rightarrow0$). In this case
the equation relates a system at late times $t$ to the density of a system with 
a large initial density $\lt{\varrho_0}$ measured at early times $\lt{t}$.
%
%
%
%
%
%
%
%
%
%
\nsubsecnn{The late-time density excess below two dimensions}
The scaling equation derived above is now used to obtain the late-time
 density. Using the notation for early-time quantities defined above, the perturbative excess, Eq. (\ref{PE}) is rewritten with $t\rightarrow\wt{t}$ etc and inserted into the scaling equation (\ref{denflow}). In dimensions $d=2-\epsilon$ the quantity $\wt{g}$ tends to its fixed point value $\wt{g}=2\pi\e$ as the ratio $t/\wt{t}$ increases. The spatially independent and dependent density components become in the late-time limit
\begin{eqnarray}
\varrho_B(t)&=&\frac{1}{(Dt)^{d/2}}\left[\frac{1}{4\pi\e}+\frac{2\log(8\pi)-5}{16\pi}\right]+O(\e), \\
\varrho_E(z,t)&=& \frac{1}{8\pi(Dt)^{d/2}}f_2\left(\frac{z^2}{2Dt}\right)+O(\e).
\end{eqnarray}
As expected, the bulk value is unchanged from the well known case derived in \cite{BPL} and given in Table \ref{bulktable}. However, also present is an excess $\varrho_E$ that decays at the same rate as the bulk case, i.e. it has a fixed amplitude with respect to $\varrho_B$. It shares the same universality as the bulk case in as much as it is independent of $\varrho_0$ and the reaction rate $\lambda$. It should be noted that the excess is not localized at the boundary but extends diffusively into the system by virtue of the functional dependence of $f_2$ on the ratio $z^2/Dt$.

\nsubsecnn{The late-time density excess in two dimensions}
In two dimensions the beta function takes a different form $\beta_g\propto g^2$. In this case we have the scaling of the reaction parameter in the perturbative density $\wt{g}\sim[4\pi\log(t/\wt{t})]^{-1}$. The leading order terms in time are now taken, which involves the RG-improved tree-level calculation for the bulk density and the one-loop calculation for the excess density
\begin{eqnarray}
\varrho_B(t)&=&\frac{\log(Dt/D\wt{t})}{8\pi Dt}+O\left(\frac{1}{t}\right)\;\;=\;\;\frac{\log(t)}{8\pi Dt}+O\left(\frac{1}{t}\right) \nonumber \\
\varrho_E(z,t)&=& \frac{1}{8\pi Dt}f_2\left(\frac{z^2}{2Dt}\right)+O\left(\frac{1}{t\log(t)}\right). \label{2d2}
\end{eqnarray}
Again, the bulk density is the known result given in Table \ref{bulktable}. Now the leading-order term of the excess is the fully universal, asymptotically exact result for two dimensions. We will discuss this further in the summary.

\subsection{Exact solution in one-dimension}
The renormalization group has provided the universal properties of the density excess as a function of dimension and the late-time form was found exactly for the case of two dimensions. However, the epsilon expansion is not expected to give accurate results in one dimension. In this section we solve the one-dimensional case with an infinite reaction rate. However, by virtue of the universality demonstrated above, the results to be described are also valid for any finite reaction rate. 

The model is now defined in dimension $d=1$ with an infinite on-site reaction rate.
Since the reaction rate is infinite, each lattice 
site can be occupied by at most one particle. Denoting a particle on site $k$ by $A_k$ and 
an empty site by $O_k$ the dynamics of the model are then
\begin{eqnarray}
O_kA_{k+1}&\leftrightarrow&A_kO_{k+1} \mbox{ with a rate D} \nonumber \\
A_kA_{k+1}&\rightarrow&O_kO_{k+1} \mbox{ with a rate 2D},  \label{rates}
\end{eqnarray}
where $k$ is restricted to the positive integers. At $t=0$ the state of the system
is chosen to be an uncorrelated random initial state with density $1/2$.

Homogeneous reaction systems in one dimension have been solved exactly by various 
methods \cite{KANGRED}-\cite{LUSH}. For inhomogeneous systems, a general method was
 recently proposed
in which the model is solved by mapping it onto a dual system \cite{SCHUTZ,SCHUTZ2}.
 Using
this method, the master equation is first written in 
a quantum spin chain representation. Each configuration is represented by a vector
 $\ket{s_1,s_2,\ldots } \equiv \ket{\lbrace s_i \rbrace}$, where $s_i$ takes the value $1/2$
 if site $i$ is empty and 
$-1/2$ is site $i$ is occupied. The state ket of the system is defined similarly to the
 bosonic case to be 
\begin{eqnarray}
\ket{P(t)}&=&\sum_{\lbrace s \rbrace} P( \lbrace s_i \rbrace : t)\ket{\lbrace s_i \rbrace}\hspace{1cm}\partial_t|P\!>\;\;=\;\;-{\cal H}|P\!> \nonumber \\
{\cal H}&=&-D\sum_{k=1}^{\infty}\left( s_k^+ s_{k+1}^- +s_{k+1}^+ s_{k}^- + 2s_k^+ s_{k+1}^+ - (1- \sigma_k^z)/2 \right). 
\end{eqnarray}
Here $s_k^{\pm}=(\sigma_k^x \pm i \sigma_k^y)/2$, where the $\sigma$'s are Pauli 
matrices. Averages are calculated in the same way as the bosonic case, by using the projection state $\bra{}=\sum_{\lbrace s \rbrace} \bra{\lbrace s_i \rbrace}$. It was shown that using a similarity transformation the system can be mapped onto a dual process with the following dynamics 
\begin{eqnarray}
A_kA_{k+1}&\rightarrow&0_k0_{k+1} \hspace{1.1cm} \mbox{ rate 2D for sites $k=1\cdots \infty$} \nonumber \\
0_kA_{k+1}&\rightarrow&A_k0_{k+1} \hspace{1cm}\mbox{ rate D for sites $k=0\cdots \infty$} \label{dynaexact} \\
A_k0_{k+1}&\rightarrow&0_kA_{k+1} \hspace{1cm}\mbox{ rate D for sites $k=1\cdots\infty$} \nonumber \\
A_1A_2&\rightarrow&0_10_{2} \hspace{1.5cm} \mbox{ rate D}. \nonumber 
\end{eqnarray}
Denoting the dual Hamiltonian as $\hat{{\cal H}}$, the required density takes the form
$$
\varrho_k(t)=\frac{1}{2}\Ex{2|e^{-\hat{{\cal H}}t}|k-1,k}.
$$ 
Here $\bra{2}$ is the sum of all two-particles states and $\ket{k-1,k}$ is the initial state with only two particles located at sites $k$ and $k-1$. However, since the reaction
rate is infinite the two-particle
transition probability can be written in terms of single-particle transition probabilities, giving the density as
\begin{equation}
\varrho_k(t)=\frac{1}{2}\sum_{m=0}^{\infty}\sum_{n=m+1}^\infty\left[T_{m,k-1}(t)T_{n,k}(t)-T_{n,k-1}(t)T_{m,k}(t)\right],
\label{exactden}
\end{equation}
where $T_{n,k}(t)$ is the probability of a particle starting at site $n$ to be at site $k$
 at time $t$ in the dual system with the forms
\begin{eqnarray}
T_{z,a}(t)&=&B(z-a : t)+B(z+a : t) \nonumber \\
T_{0,a}(t)&=&1-\sum_{m=1}^\infty T_{m,a}(t). \nonumber
\end{eqnarray}
Here $B(m : t)=e^{-2Dt}I_{m}(2Dt)$ is the single-particle propagator with $I_m$ the order-$m$ modified Bessel function. Using these, results the double sum (\ref{exactden}) is reduced to
\begin{eqnarray}
\varrho_k(t)&=& \frac{1}{2} \Big[-B(1-k : t)\sum_{m=3}^\infty B(m-k : t)+\sum_{m=2}^\infty B(m-k : t)\left[B(m-k : t)+B(m+1-k : t)\right]\nonumber \\
&&+ B(k : t)\sum_{m=2}^\infty B(m+k : t)-\sum_{m=1}^\infty B(m+k :t )\left[B(m+k:t)+B(m+1+k:t)\right]\nonumber \\
&& -B(k:t)\sum_{m=2}^\infty B(m-k:t)+ B(1-k:t)\sum_{m=1}^\infty B(m+k:t)\nonumber+B(k:t)+B(1-k:t)\Big]. 
\end{eqnarray}
Taking the late-time limit, which is valid when $\sqrt{Dt}$ is much greater than the lattice spacing and writing $k=z$, we obtain
\begin{eqnarray}
\varrho(z,t)&=& \frac{1}{\sqrt{8\pi Dt}}\left[ \mbox{erf}\left(\frac{k}{\sqrt{2Dt}}\right)+\sqrt{2}\mbox{erfc}\left(\frac{k}{\sqrt{4Dt}}\right)\exp\left(-\frac{k^2}{4Dt}\right)\right]\nonumber\\
\varrho(z,t)&=&\frac{1}{\sqrt{8 \pi Dt}}+\varrho_E(z,t),
\end{eqnarray}
where again the density is split into bulk and excess $\varrho_E$ components
\begin{eqnarray}
\varrho_E(z,t)&=&\frac{1}{\sqrt{8\pi Dt}}f_1\left(\frac{z^2}{2Dt}\right), \label{1dexact} \\
f_1(\xi^2)&=&\sqrt{2}\mbox{erfc}\left(\frac{\xi}{\sqrt{2}}\right)\exp\left(\frac{-\xi^2}{2}\right) - \mbox{erfc}\left(\xi\right).\nonumber
\end{eqnarray}
As expected the bulk component of the density
 $\varrho_B$ is found to be identical to the infinite, unbounded case \cite{LUSH0,LUSH}. The asymptotic form of the function is $f_1(\xi^2) \sim \exp(-\xi^2/2)/\xi$.

\subsection{Summary of the diffusive case}
\begin{figure}[t]
\hspace{2.5cm}
\epsfxsize 11 cm
\epsfysize 10 cm
\epsfbox{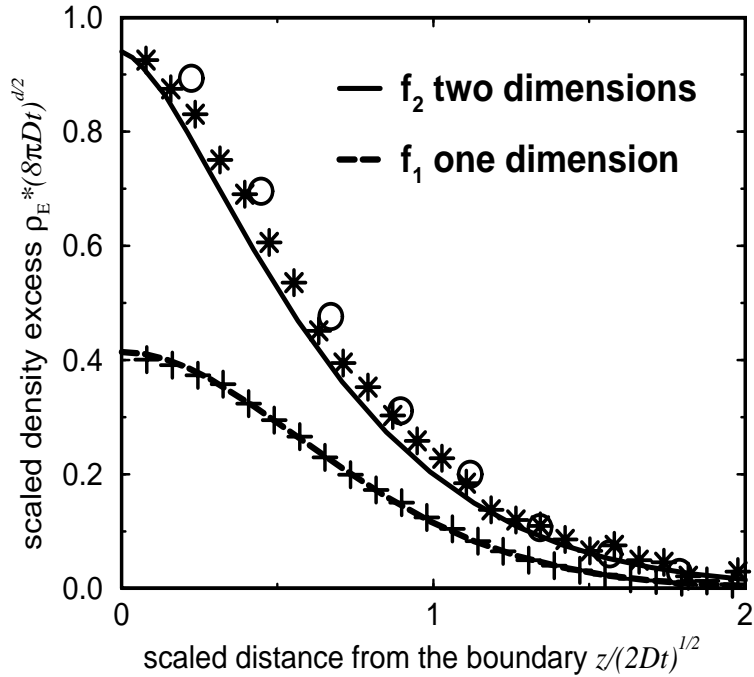}
\caption{The density excess in one and two dimensions, $f_1$ and $f_2$. The 
renormalization group results are compared with simulations at time $t=10$ ($\circ$) and $t=80$ ($*$). The exact results $f_1$ calculated for infinite reaction rate is compared with simulations for finite reaction rate $\lambda =1/2$ ($+$) demonstrating the universality.}
\label{2dfig}
\end{figure}

Results have been presented from an analysis of a reaction-diffusion process
 near an impenetrable boundary. It was found that in two dimensions and below a density excess forms near the boundary and extends into the bulk diffusively. The density can be written in the following form
$$
\varrho(z,t)=\varrho_B(t)+\varrho_E(z,t), \nonumber
$$
where $\varrho_B(t)$ is the bulk result seen in a translationally invariant system and $\varrho_E$ is the boundary-induced excess density - see Table \ref{bulktable}. Using the field-theoretic renormalization group the universal properties of the density excess were found. It was shown that the excess is independent of both the initial density and the reaction rate $\lambda$. In two dimensions it was found that the density excess is marginally less dominant than the bulk, with the following asymptotic (large $z^2/2Dt$) limit
\begin{eqnarray}
\varrho_E&=&\frac{1}{8\pi Dt}\left(\frac{2Dt}{z^2}\right)^{3/2}\exp\left(-\frac{z^2}{4Dt}\right)\mbox{ in two dimensions}.
\end{eqnarray}
The full asymptotically-exact late-time result is given by (\ref{2d2}) with the double integral (\ref{2d}) and is plotted in Fig. \ref{2dfig}.
\begin{table}[t]
\begin{center} 
\leavevmode
\begin{tabular}{|c|c|c|}
\hline
$d$ & late-time bulk density $\varrho_B$ & late-time excess density $\varrho_E$\\
\hline
\hline
$1$ & $\frac{1}{(8 \pi Dt)^{1/2}}$&$\frac{1}{(8\pi Dt)^{1/2}}f_1\left(\frac{z^2}{Dt}\right)$ \\
\hline
$2-\epsilon$ & $\frac{1}{4 \pi \epsilon (Dt)^{d/2}} \left[1+ \frac{\epsilon}{4} \left( 2 \log(8 \pi) -5 \right) \right]+O(\epsilon)$ &$\frac{1}{(8\pi Dt)^{d/2}}f_2\left(\frac{z^2}{Dt}\right)+O(\epsilon)$\\
\hline
$2$ & $\frac{\log(t)}{8 \pi Dt}$& $\frac{1}{(8\pi Dt)}f_2\left(\frac{z^2}{Dt}\right)$\\
\hline
$>2$ & $\frac{1}{2 \lambda t}$& sub-leading\\
\hline
\end{tabular}
\caption{The known results for the late-time bulk density and the new results for the bounded system.}
\label{bulktable}
\end{center}
\end{table}

In one dimensions the excess density was obtained exactly (\ref{1dexact}) by the solution of a corresponding quantum spin system and is also plotted in Fig. \ref{2dfig}, with the asymptotic (large $z^2/2Dt$) limit
\begin{eqnarray}
\varrho_E&=&\frac{1}{(8\pi Dt)^{1/2}}\left(\frac{2Dt}{z^2}\right)^{1/2}\exp\left(-\frac{z^2}{4Dt}\right)\mbox{ in one dimension}.
\end{eqnarray}
Here we see that the excess and bulk density share the same leading order decay of $\varrho\sim t^{-1/2}$. The ratio between the density at the boundary and the density far in to the bulk below two dimensions is thus independent of all system parameters except the dimension of space. In one dimension it is numerically equal to $\sqrt{2}$ and it is interesting to note that the renormalization group gives quite a fair indication of this ratio
\begin{equation}
\frac{\varrho(0,t)}{\varrho_B(t)}=1+\epsilon\left(\frac{3}{4}+\frac{\pi}{2}-\frac{3\pi^2}{16}\right)+O(\epsilon^2), \label{ratio}
\end{equation}
which is $\sim1.47$ for one dimension ($\epsilon=1$). It should also be noted that above the critical dimension there is a sub-dominant
density excess. However, these effects are transient and quickly decay to
yield the predicted mean-field result.

We end the discussion of the diffusive case by considering the implications
 of the solution in $d=1$ to the coarsening dynamics of the critical Ising
 model. As explained earlier, the domain wall dynamics in this Ising system can be mapped onto the mutually annihilating random walk \cite{ISING}. The boundary in the reaction system corresponds to a fixed spin in the magnetic system. Consider the magnitude of 
the coarse-grained magnetization near this fixed spin. This quantity is a function of the local density of domain
walls - the fewer the domain walls the higher the magnetization. The density excess (\ref{1dexact}) implies a higher density of domain walls near the fixed spin. This yields the counter-intuitive result that the absolute value of the coarse-grained magnetization is {\it lower} near fixed spin.

\newpage

\section{Ballistic annihilation near a boundary in one dimension}
In the previous section the effect of an impenetrable boundary on
the single-particle reaction-diffusion process $A+A \rightarrow 0$ was examined. It was found that for diffusing reactants a density excess was produced at the boundary. In this section we will demonstrate that this effect is not independent of the reactants' dynamics: for the case of ballistic annihilation a density {\it deficit} is formed.

By ballistic motion, it is meant that reactants travel in straight lines until they collide and a reaction occurs. These dynamics are relevant for gas-phase and some excitonic reactions \cite{BUT} and a good review can be found in \cite{PRIV}. Because of the deterministic dynamics of individual reactants, the only sources of noise are in the initial distribution of particles or if the reaction rate is finite. Though a great deal of analytic work exists on these systems, they have largely examined two kinds of initial conditions in one dimension. One form of initial condition was introduced \cite{PRD} to examine the dynamics of reaction fronts and has been demonstrated to be universal with respect to a finite reaction rate \cite{MJER}. However, the case that will interest us in this section is that of random initial conditions whereby particles are distributed upon a line with each particles' velocity drawn from the same distribution. A broad variety of initial-velocity distributions have been studied \cite{PRIV,DRFP,KRL} and universal effects analyzed for the continuous case \cite{RDP}. However, in these cases the systems that have been examined have been infinite and translationally invariant.

In this section we will generalize the binary-velocity model introduced by Elskens and Frisch \cite{EF}. The behavior of this translationally invariant system will be recalled below and the generalization to the case of a semi-infinite system bounded by a reflecting wall will be introduced. The basic results will be stated and then details of the method given. Finally, the section is closed with a summary of the behavior in the late-time limit.

\nsubsecnn{Summary of known results for the unbounded system}
The model introduced in \cite{EF} comprises an infinite line upon which particles move with fixed velocity $+c$ or $-c$. When two particles meet a mutual annihilation always occurs, and both particles are removed from the system. The initial conditions provide the only source of noise and are defined such that at time $t=0$ particles are equally spaced along the line and given either of the two fixed velocities with equal probability. A mean-field approach predicts that the density decay should vary as $\varrho\sim t^{-1}$. Nevertheless, the density decay derived from a full solution of the model has the form $\varrho\sim t^{-1/2}$. This is the same exponent as for the diffusive case, but it should be stressed that whereas in the diffusive case the changed exponent comes from a dynamic effect, in ballistic annihilation it is a property only of the initial conditions. This can be seen from the following qualitative argument. Consider the density fluctuations in a domain of length $\ell$. The typical fluctuation in the difference of particle numbers with $v=+c$ and $v=-c$ inside this 
domain is $\propto \ell^{1/2}$. After a time $\propto \ell$ only 
the residual fluctuation will remain, so that the density $\varrho$ will
be $\propto \ell^{-1/2}$. Re-expressing $\ell$ as a function of $t$ one
 obtains the correct density-decay exponent $\varrho \propto t^{-1/2}$.
 
\begin{figure}
\epsfxsize 10 cm
\epsfysize 6 cm
\hspace{2.5cm}\epsfbox{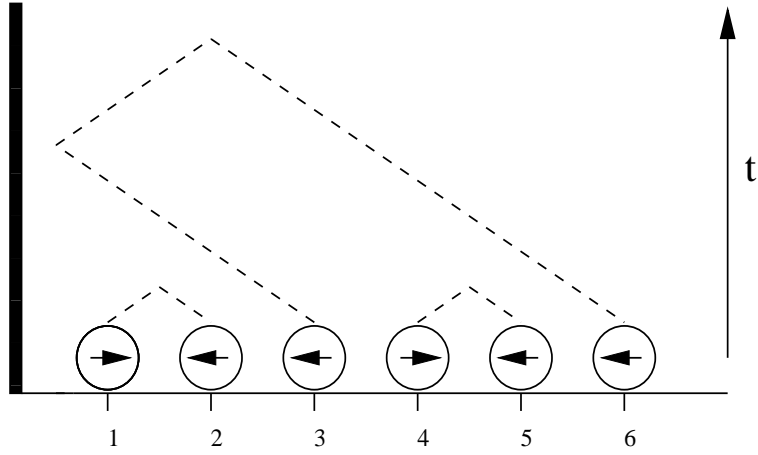}
\caption{An example of the initial conditions studied. The numbers 
mark lattice sites, and dashed lines
represent the trajectories of the particles.} 
\label{BALMODEL}
\end{figure}

\nsubsecnn{Annihilations near a boundary - new results}
A reflecting boundary is now introduced into the model defined above. The boundary is placed at position $z=1/2$ on the one-dimensional line and particles are placed, equally spaced at positions $z=1,2,\cdots$. When referring to particle $k$ it is meant that this is the particle that was at position $z=k$ at time $t=0$. Each particle is given a velocity of either $+c$ or $-c$ with equal probability. The velocity of each particle is fixed unless it reaches the impenetrable, reflecting boundary. In this case a particle with velocity $-c$ rebounds with a fixed velocity $+c$. Reactions occur with a probability of unity on contact, so that once the trajectories of two particles meet a mutual annihilation always occurs, see Fig. \ref{BALMODEL} for an illustration of the dynamics.

The quantity of interest is the density $\varrho$ averaged over all possible initial velocity distributions. It will be demonstrated below that this density can be written as a sum of a bulk, background density $\varrho_B$ equivalent to that calculated in \cite{EF} and a boundary-induced density deficit $\varrho_D$
\begin{eqnarray}
\varrho(z,t)&=&\varrho_B(t)+\varrho_D(z,t).\nonumber
\end{eqnarray}
The scaling form of the density deficit, given exactly in the following sub-sections, is 
\begin{eqnarray}
\varrho_D(z,t) & = & - \frac{1}{(ct)^{1/2}}g \left( \frac{z}{ct} \right), \label{BALSCALING}
\end{eqnarray}
and should be compared to the diffusive form which decays with the same exponent $\sim t^{-1/2}$ but is a function of the dimensionless quantity $z^2/Dt$. In the ballistic case it is also seen that the effect penetrates from the boundary into the bulk, though here it is ballistic by virtue of the dependence of $g$ on the dimensionless form $z/ct$.

A detailed description of the exact solution now follows. First, the probability of any specific pair of particles to react is calculated. The method of solution is combinatoric and related to the well-known {\it ballot problem} \cite{FELLER} of probability theory. From this fundamental pair-reaction probability all other quantities are obtained, including the density profile and the profiles for left and right moving particles. 

\subsection{Exact solution in one-dimension}

\nsubsecnn{The pair annihilation probability}
Because the model defined above has a reaction probability of one, only mutual annihilations between even and odd particles are possible. There are two  cases that need to be considered: either the even or the odd particle can be nearest the boundary. We obtain all the possible pair-annihilation probabilities by considering an odd numbered particle $k$ annihilating either with even particle $k+(2n+1)$
 or even particle $k-(2n+1)$ where $n=0,1,\cdots$.  

Consider first the probability of the odd numbered particle $k$ to mutually annihilate with particle $k+(2n+1)$. This probability has two contributions: (i) direct annihilation with particle $k$ having initial velocity $v=+c$ and (ii) indirect annihilation with particle $k$ having an initial velocity $v=-c$ but rebounding from the wall. For direct annihilation the calculation is identical to the homogeneous case \cite{EF}. A mapping exists to the problem of a random walker returning to the origin after $2n+2$ steps for the first time - this is the so-called ballot problem. The probability for this to happen, given that the initial velocity of particle $k$ is $+c$ is
\begin{equation}
\frac{1}{2n+1}\Big(\frac{1}{2}\Big)^{2n+1}{2n+1 \choose n+1}.
\label{HOMOG}
\end{equation}
\begin{figure}
\epsfxsize 10 cm
\epsfysize 6 cm
\hspace{2.5cm}\epsfbox{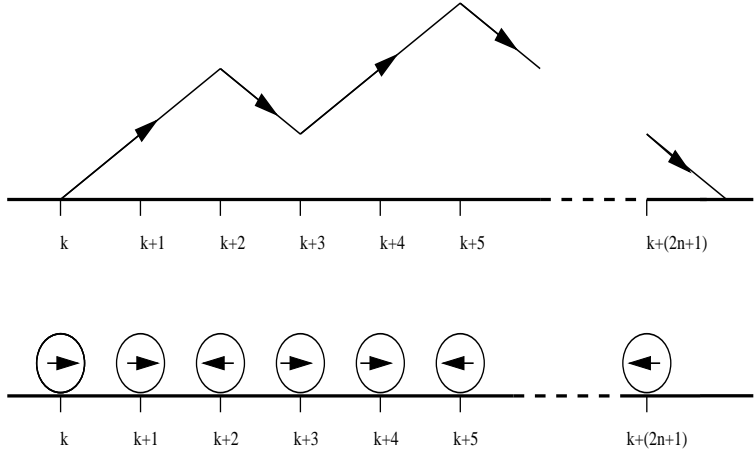}
\caption{An illustration of the mapping to the ballot problem. The arrows,
 representing the direction a particle is
 moving, are considered as moves of a random walker as illustrated above.} 
\label{ballot}
\end{figure}
Next, consider the contribution from the indirect case whereby a particle is first reflected from the boundary before annihilating. In this case the initial velocity of particle $k$ is $v=-c$ with the total probability given by the product of Eq. (\ref{HOMOG}) and the probability that particle $k$ reaches the boundary. The latter can be found by noting that particles labeled by an
 even index
 with an initial velocity $v=-c$ are always annihilated before
 arriving at the wall. It is therefore equivalent to the probability of a random walker not to have returned to the origin after $k-1$ steps,
\begin{equation}
\Big(\frac{1}{2}\Big)^{k-2} \sum_{x=2,4 \ldots}^{k-1} \frac{x}{k-1} 
{k-1 \choose \frac{x+k-1}{2}}=\Big(\frac{1}{2}\Big)^{k-2}{k-2 \choose \frac{k-3}{2}}
\label{WALLCON}
\end{equation}
where this is conditional on particle $k$ having an initial velocity $v=-c$. Combining the two results of Eqs. (\ref{HOMOG}) and (\ref{WALLCON}) gives the net probability that particle $k$ interacts with particle $k+(2n+1)$, $P_{k,k+(2n+1)}$ regardless of particle $k$s initial velocity
\begin{equation}
P_{k,k+(2n+1)}=\frac{1}{2n+1}\Big(\frac{1}{2}\Big)^{2n+2}{2n+1 \choose n+1}\left(1+
\Big(\frac{1}{2}\Big)^{k-2}{k-2 \choose \frac{k-3}{2}} \right).
\label{rightcon}
\end{equation}
The final case of the odd numbered particle $k$ to interact with particle $k-(2n+1)$, $P_{k,k-(2n+1)}$, can also be derived using the arguments given above
\begin{equation}
P_{k,k-(2n+1)}=\frac{1}{2n+1}\Big(\frac{1}{2}\Big)^{2n+1}{2n+1 \choose n+1}.
\label{leftcon}
\end{equation}

The Eqs. (\ref{rightcon}) and (\ref{leftcon}) give all the possible pair annihilation probabilities
in the system, where it should be remembered that particle $k$ is odd numbered. 

\nsubsecnn{The density calculation}
To ease the notation in this section we set $c=1$, corresponding to a trivial redefinition of time. The initial conditions of equally spaced particles introduces a discretization of time into the problem. We consider the model at integer time steps when particles will be found only at positions $z=1,2\cdots$ which we call lattice sites. In the calculation of the density two cases must be considered: (i) $z \geq t$ when the presence of the boundary does not affect the
 density profile, (ii) $z<t$ where the boundary does affect the profile.\\

\noindent {\it (i) The case  $z \geq t$}\\ 
Here the density is the same 
as for the infinite-system case \cite{EF}. The contribution
to $\varrho(z,t)$ can arise from particle $z+t$ and $z-t$. Using the mapping
to the ballot problem we find
\begin{eqnarray}
\varrho(z,t)& = &\Big( \frac{1}{2} \Big)^{2t-1} {2t-1 \choose t-1},
\label{BALBULK}
\end{eqnarray}
which is the probability of particle $z-t$ not to interact with particles up to and including $z+(2t-1)$ added to an equal contribution from particle $z+t$.
Note that the results are valid for $t=z$ since the contribution from a particle initially at
site $1$ with velocity $v=-1$ which rebounded off the wall is equivalent to the contribution from a ``ghost'' particle initially at site $0$ with velocity $+1$.\\

\begin{figure}
\centerline{
	\psfig{figure=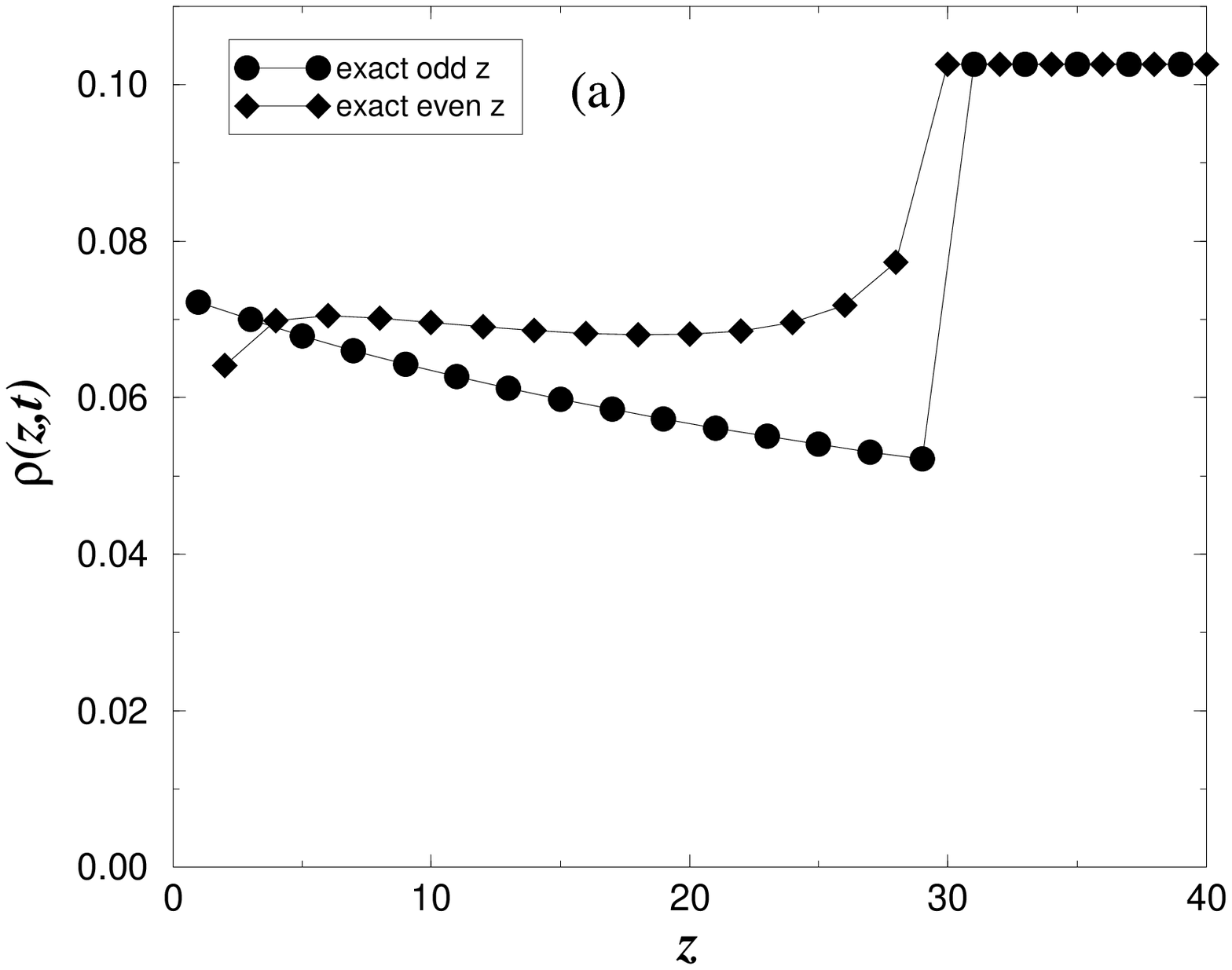,width=8cm}
	\psfig{figure=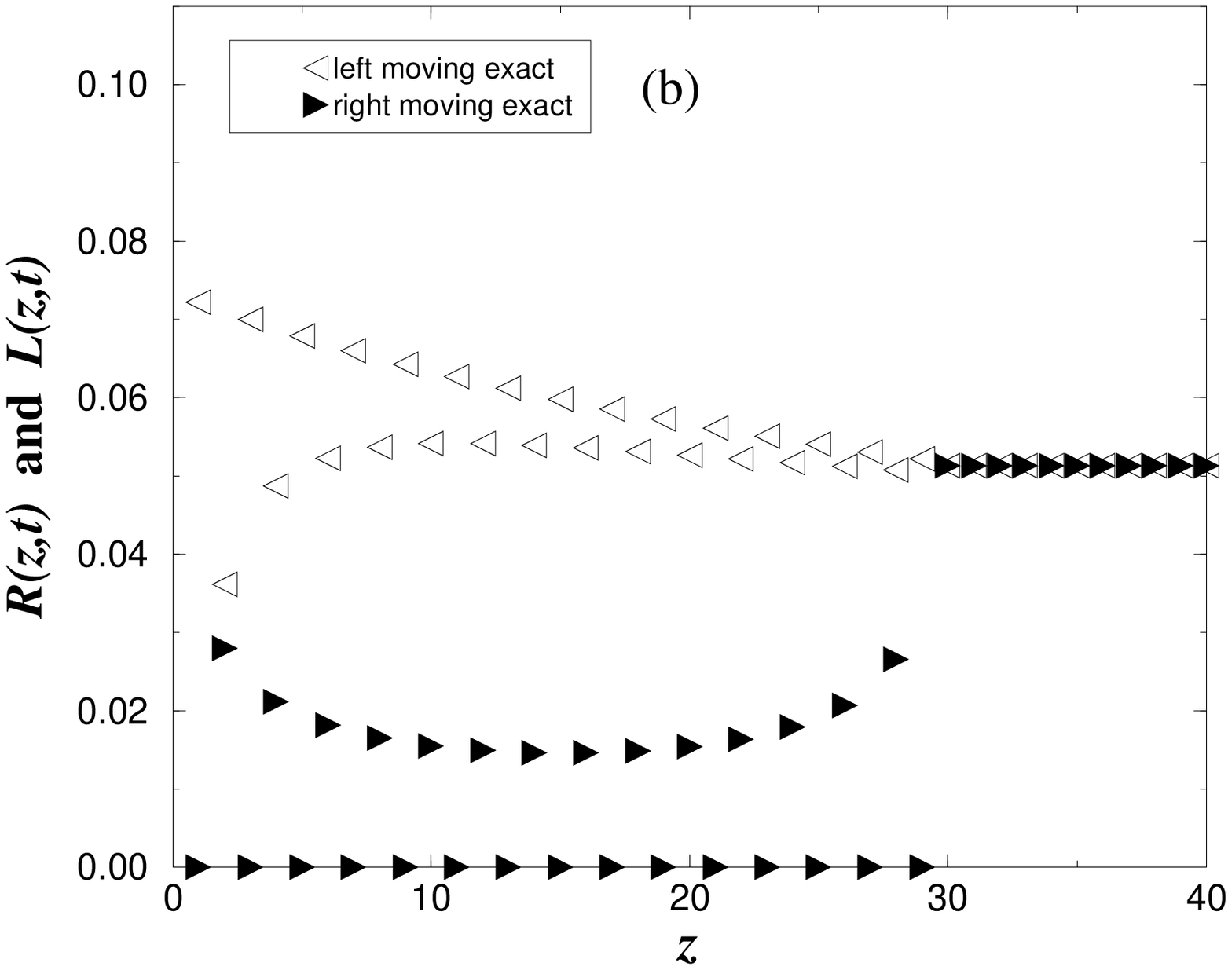,width=8cm}
}
\caption{An example of the exact (a) density profile and (b) densities of left and right moving particles at time $t=30$. The difference between odd and even sites becomes increasingly less pronounced at later times.}
\label{exactb1}
\end{figure}

\noindent{\it (ii) The case $z>t$}\\
There are two contributions 
to the density at site $z$, one from particle $z+t$
 with an initial velocity $v=-1$ and another from particle $t-z+1$ with an
 initial velocity $v=-1$
 which rebounded from the wall at time $t-z+1/2$.
 First, the contribution
from particle $z+t$ is considered with $z$ odd. This particle reaches site $z$ 
if it does not annihilate with any of the $z+t-1$ particles to its left. 
When $z+t$ is odd ($t$ is even) we obtain for this probability
\begin{eqnarray}
\Big(\frac{1}{2}\Big)^{z+t-1}\sum_{x=2,4, \ldots}^{z+t-1}\frac{x}{z+t-1}
{z+t-1 \choose \frac{z+t-1+x}{2}}&=& \Big( \frac{1}{2} \Big)^{z+t-1} { z+t-2 \choose \frac{z+t-3}{2}}.
\nonumber
\label{ioddteven}
\end{eqnarray}
When $z+t$ is even ($t$ is odd) the particle never reaches the wall so that the
 probability of a particle initially at 
site $z+t$ of surviving up to site $z$ is the sum of all the chances of being
 annihilated after site $z$
\begin{eqnarray}
& &\sum_{x=1,3, \ldots}^{2z-3} \Big( \frac{1}{2} \Big)^{z+t-1} \frac{1}{x} { z+t-x-2 \choose
 \frac{z+t-x-3}{2}}
{x \choose \frac{x+1}{2}} \nonumber\\
& =& \Big( \frac{1}{2} \Big)^{z+t-1} \left[ \frac{z+t-1}{z+t}{z+t-2 \choose \frac{z+t-2}{2}} -
 \frac{t-z+1}{z+t}
{ t-z \choose \frac{t-z}{2}}{2z-2 \choose z-1} \right]. \nonumber \label{ioddtodd}
\end{eqnarray}
Now, the contribution from particle $t-z+1$ is considered also with $z$ odd. This is non-zero only
if $t-z+1$ is odd ($t$ is odd), and is then given by,
\begin{eqnarray}
& &\Big(\frac{1}{2}\Big)^{t-z}{t-z-1 \choose \frac{t-z-2}{2}} \Big(\frac{1}{2}\Big)^{2z-1}
\sum_{x=2,4, \ldots}^{2z} \frac{x}{2z}{2z  \choose \frac{2z+x}{2}} \nonumber\\
&=& \Big(\frac{1}{2}\Big)^{t+z-1}{t-z-1 \choose \frac{t-z-2}{2}} {2z-1 \choose z-1}. \nonumber \label{ioddtoddb}
\end{eqnarray}
Combining the above results from particle $z+t$ and particle $t-z+1$ we
 obtain the density at odd sites $z$ and at even and odd times. Using similar argumentation allows the results for even sites also to be found. The forms for the densities are
\begin{eqnarray}
\varrho(z_{odd},t_{even})&=&\varrho(z_{even},t_{odd})=\Big( \frac{1}{2} \Big)^{z+t-1} { z+t-2 \choose \frac{z+t-3}{2}}, \nonumber \\
\varrho(z_{odd},t_{odd})=\varrho(z_{even},t_{even})\!\!&=&\!\! \Big( \frac{1}{2} \Big)^{z+t-1} \Big[ \frac{z+t-1}{z+t}{z+t-2 \choose \frac{z+t-2}{2}} - \frac{t-z+1}{z+t}{ t-z \choose \frac{t-z}{2}}{2z-2 \choose z-1} \nonumber\\
& &+{t-z-1 \choose 
\frac{t-z-2}{2}} {2z-1 \choose z-1} \Big]. \nonumber 
\end{eqnarray}\\
These results taken together represent the complete density profile at all times. An example is shown graphically in Fig. \ref{exactb1}. It is clear from these results that a density deficit is indeed formed at the boundary. It should also be commented upon that the even/odd effect, though strong at early times, becomes increasingly less significant at later times. We now consider how this density is distributed between left and right moving particles.\\

\nsubsecnn{The density of left and right moving particles}
For the case of the unbounded system the density of left and right moving particles is equal by symmetry. For this reason it is only necessary to consider the case $z<t$ because the boundary has no influence for $z \geq t$. Using similar method to the above, the following forms are found\\

\noindent {\it The density of left-moving particles}\\
\begin{eqnarray}
{\cal L}(z_{even},t_{odd})&=&{\cal L}(z_{odd},t_{even})=\Big( \frac{1}{2} \Big)^{z+t-1} { z+t-2 \choose \frac{z+t-3}{2}} \label{left1}, \nonumber \\
{\cal L}(z_{even},t_{even})&=&{\cal L}(z_{odd},t_{odd})=\\ \nonumber 
&&\Big( \frac{1}{2} \Big)^{z+t-1} \left[ \frac{z+t-1}{z+t}{z+t-2 \choose \frac{z+t-2}{2}} -
 \frac{t-z+1}{z+t}
{ t-z \choose \frac{t-z}{2}}{2z-2 \choose z-1} \right]. \nonumber 
\end{eqnarray}\\

\noindent {\it The density of right-moving particles}\\
\begin{eqnarray}
{\cal R}(z_{odd},t_{even})&=&{\cal R}(z_{even},t_{odd})=0,\nonumber \\
{\cal R}(z_{odd},t_{odd})&=&{\cal R}(z_{even},t_{even})=\Big(\frac{1}{2}\Big)^{t+z-1}{t-z-1 \choose \frac{t-z-2}{2}} {2z-1 \choose z-1}. \nonumber
\end{eqnarray}\\
As an example, the left and right densities for the even time of $t=30$ are plotted in the second graph of Fig. \ref{exactb1}. From this graph it can be clearly seen why the deficit occurs. As time progresses the remaining particles tend to congregate in groups which move in the same direction. Once such a group hits the boundary it mostly annihilates within itself, 
reducing the density of returning (right-moving) particles. 

\subsection{Summary of ballistic case}
\begin{figure}
\epsfxsize 8 cm
\epsfysize 7 cm
\hspace{4cm}\epsfbox{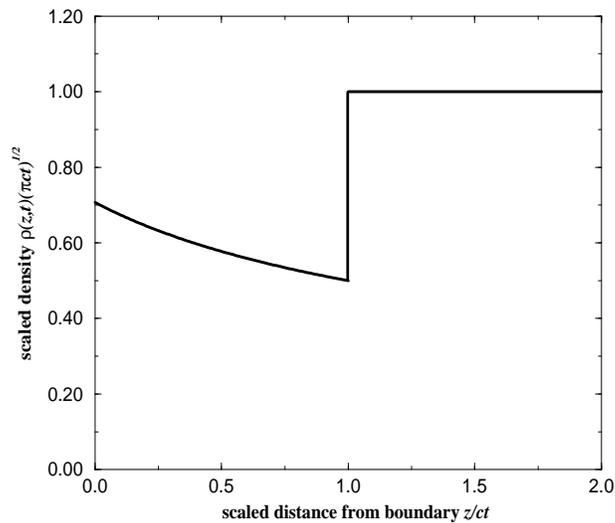}
\caption{The late-time scaled density profile near the boundary.} 
\label{scaledbal}
\end{figure}

\begin{figure}
\centerline{
	\psfig{figure=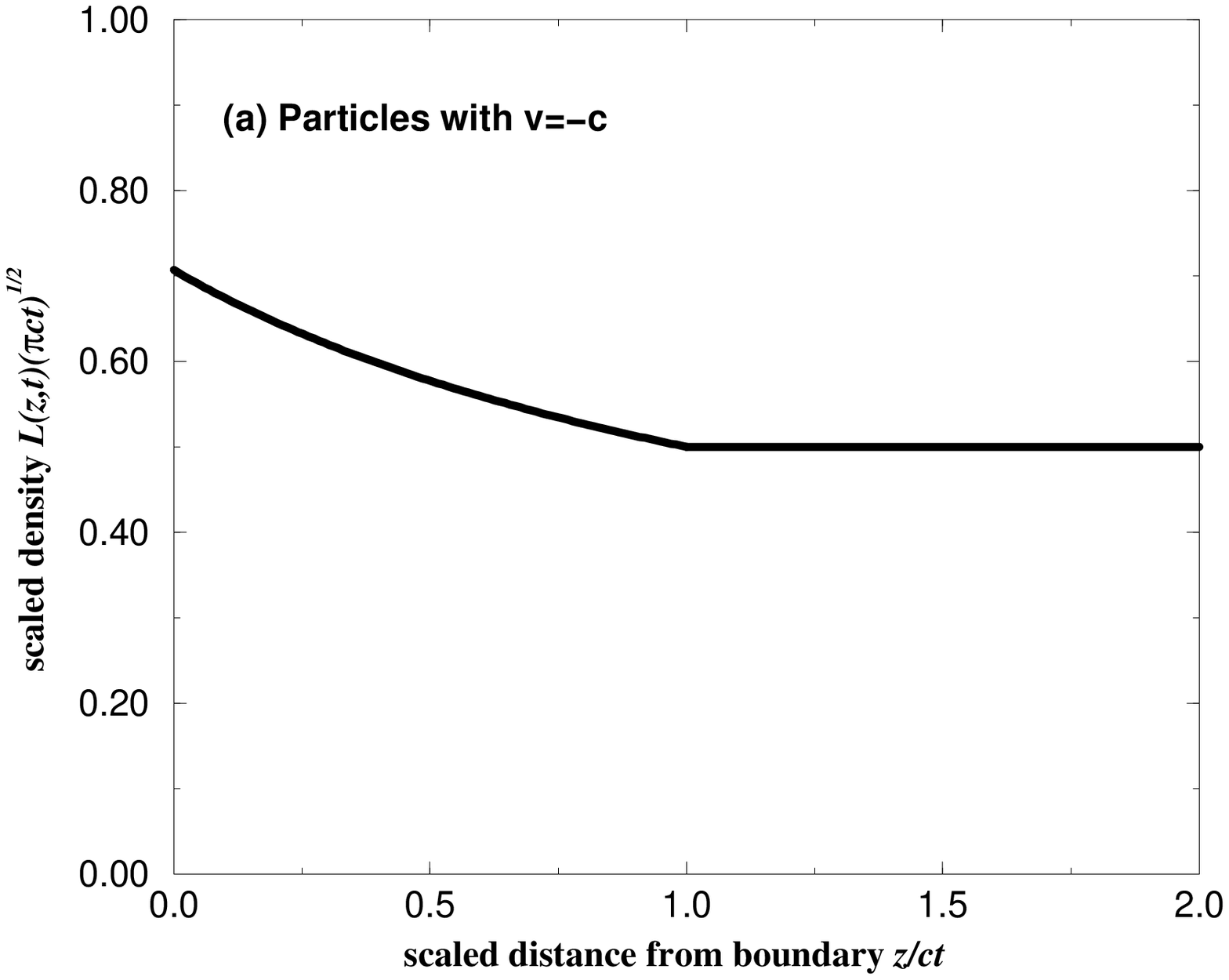,width=8cm}
	\psfig{figure=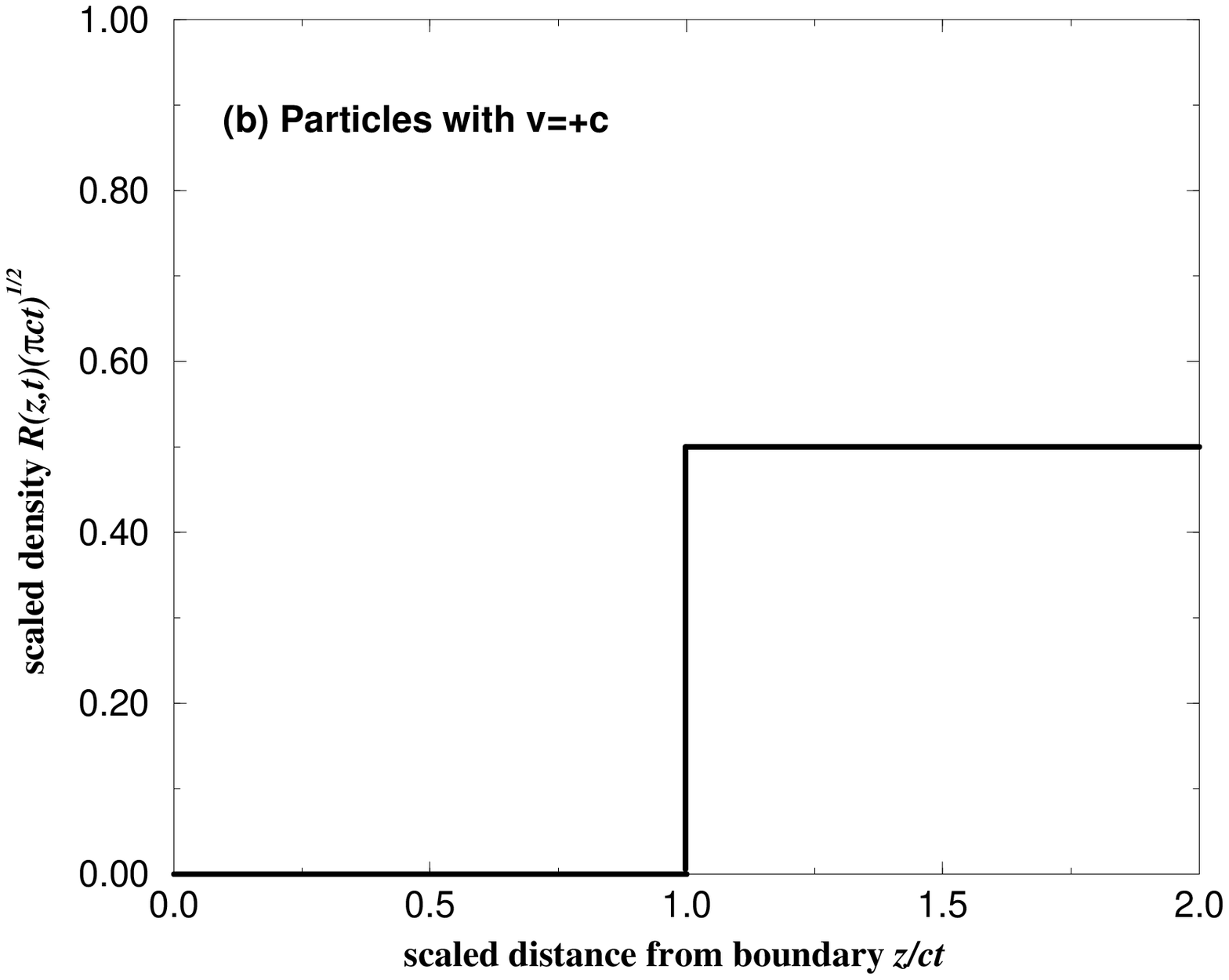,width=8cm}
}
\caption{The late-time scaled density of (a) left-moving particles with $v=-c$ and (b) right-moving particles with $v=+c$.}
\label{scalerlbal}
\end{figure}

The effect of an impenetrable, reflecting boundary has been examined on a one-dimensional realization of ballistic annihilation. The system evolves deterministically from an initial random velocity distribution and various quantities, averaged over different realizations of this velocity distribution, were calculated exactly. In particular the density profile of the reactants as a function of time and distance from the boundary was obtained. It was found that, in contrast to the diffusive case studied in the previous section, a density deficit $\varrho_D(z,t)$ was formed at the boundary that extends into the system ballistically 
\begin{eqnarray}
\varrho(z,t)&=&\varrho_B(t)+\varrho_D(z,t)\nonumber \\
\varrho_D(z,t) & = & -\frac{1}{(ct)^{1/2}}g \left( \frac{z}{ct} \right). \nonumber 
\end{eqnarray}
In the late-time limit, the scaling function $g$ takes the form  given in table \ref{bal1table} and the late-time form of the full density is shown in Fig. \ref{scaledbal}. Also calculated were the density profiles for the left and right moving particles, of which the asymptotic functional forms are given in table \ref{bal1table} and are plotted in Fig. \ref{scalerlbal}. Though the specific initial conditions of equally spaced particles was considered it is expected to hold for more general particle distributions. For example, the case of Poissonianally distributed particles can be shown to give the same late-time results. From the physical argument given above for the deficit, it might be expected that similar behavior is also seen in higher dimensions. 
\begin{table}[t]
\begin{center} 
\leavevmode
\begin{tabular}{|c||c|c|c|c|}
\hline
& $\varrho_B(t)$ & $\varrho_D(z,t)$ & ${\cal L}(z,t)$ & ${\cal R}(z,t)$ \\
\hline
\hline
$z<ct$ &$\frac{1}{\sqrt{ \pi ct}}$ & $- \frac{1}{\sqrt{ \pi ct}} \left(1- \frac{1}{\sqrt{2}\sqrt{1+z/ct}} \right)$&$\frac{1}{\sqrt{2 \pi ct} \sqrt{1+z/ct}}$ &0 \\
\hline
$z\geq t$ &$\frac{1}{\sqrt{ \pi ct}}$ &0 &$\frac{1}{ \sqrt{4 \pi ct}}$ &$\frac{1}{ \sqrt{4 \pi ct}}$ \\
\hline
\end{tabular}
\caption{The late-time scaling forms for the bulk, deficit and left and right moving density profiles.}
\label{bal1table}
\end{center}
\end{table}

\newpage

\noindent{\Large{\bf Discussion}}\\[5mm]
In this paper we have considered the effect of an impenetrable boundary on the single-species reaction $A+A\rightarrow O$. The boundary breaks the translational invariance of the system and introduces a spatial dependence into the reactant density profile. Two different dynamics for the reactants were considered: diffusive and ballistic propagation. For the case of reactants that diffuse it was found that an excess density is formed, whereas for ballistic annihilation a density deficit was obtained. It is interesting, form the view-point of surface critical phenomena to compare the influence of the boundary at different length and time scales on these two reaction processes. In the system with diffusive reactants it was seen in the RG treatment that the impenetrable/reflecting boundary behaves the same way regardless of scale: the propagator is not renormalized and maintains the form (\ref{prop}). However, this is not the case for ballistic reactions. At late times the reactants tend to move in groups that share the same velocity. When such a group meets the reflecting boundary almost total internal annihilation will occur. Hence, contrary to the case of diffusive reactions, for ballistic annihilation the reflecting boundary becomes an {\it effectively absorbing} boundary in late times.

An obvious project for further study is the effect of an impenetrable boundary on the two-species reaction $A+B\rightarrow O$. Also, the universal properties of ballistic annihilation should be further explored. It is surprising to note that even in unbounded systems the problems of ballistic annihilation in higher dimensions, finite reaction rate and different initial velocity distributions have been relatively untouched. It should be possible to solve the model that was examined in section 2 of this paper with a finite reaction rate, using the same technique as \cite{MJER} and the same long-wavelength behavior is expected. More interesting would be to analyze the sensitivity of the evolution to different initial velocity distributions, particularly in higher dimensions. Finally, given the results presented in \cite{CB} it would be worth examining the effects of a rough boundary on a simple reaction process.\\[10mm]
\noindent{\Large{\bf Acknowledgments}}\\[5mm]
We would like to thank David Mukamel and Gunter Sch\"utz for useful discussions. The authors acknowledge support from the Israeli Science Foundation.

\end{document}